\def\sect{\frenchspacing Section }
\def\fig{\frenchspacing Fig. }

\documentclass[usenatbib]{mn2e} 
\usepackage{amsmath}
\usepackage{epsfig}
\usepackage{amssymb}

\title{Constraints on the topology of the Universe derived from the 7-year \emph{WMAP} data}

\author[P.~Bielewicz, A.~J.~Banday]{P.~Bielewicz$^1$ \thanks{E-mail:
    Pawel.Bielewicz@cesr.fr}, A.~J.~Banday$^1$\\ $^1$ Centre d'Etude
  Spatiale des Rayonnements, 9, av du Colonel Roche, BP 44346, 31028, Toulouse, France \\}

\begin{document}

\maketitle

\begin{abstract}
We impose constraints on the topology of the Universe determined from
a search for matched
circles in the temperature anisotropy patterns of the 7-year \emph{WMAP} data. We pay special attention to
the sensitivity of the method to residual foreground contamination of
the sky maps, and show that for a full sky estimate of the CMB signal
(the ILC map) such residuals introduce a non-negligible
effect on the statistics of matched circles. In order to reduce this effect,
we perform the analysis on maps for which the most contaminated
regions have been removed. 
A search for pairs of matched back-to-back circles in the higher
resolution \emph{WMAP} W-band map allows tighter constraints to be imposed on topology.
Our results rule out universes with
topologies that predict pairs of such circles with radii larger than
$\alpha_{\rm min} \approx 10^\circ$. This places a lower bound on the
size of the
fundamental domain for a flat universe of about 27.9 Gpc. This bound is
close to the upper limit on the size of Universe possible to detect by the
method of matched circles, i.e.\ the diameter of the observable
Universe is 28.3 Gpc. 
\end{abstract}


\section{Introduction}
According to General Relativity, the pseudo-Riemannian manifold with signature (3,1) is
a mathematical model of spacetime. The local properties of spacetime geometry are described by the 
Einstein gravitational field equations. However, they do not specify
the global spatial geometry 
of the universe, i.e.\ its topology. This can only be constrained by observations.

The concordance cosmological model assumes that the universe 
possesses a simply-connected topology, yet recently detected
anomalies on large angular scales in the 
cosmic microwave background (CMB) anisotropy 
suggest that it may be multiply-connected.
Evidence of such anomalies comes from the suppression of the quadrupole moment, alignment
of the quadrupole and octopole and asymmetry in the temperature
anisotropy observed in two hemispheres on the sky 
\citep{de Oliveira-Costa:2004, copi:2004, eriksen:2004, hansen:2004, schwarz:2004}.

The present constraints on topology were placed by studying two signatures of multi-connectedness in 
the CMB maps: the large scale damping of the power in the direction of
a small dimension of the domain, which causes a breakdown of
statistical isotropy \citep{de Oliveira-Costa:2004,kunz:2006,
  kunz:2008}, and the distribution of matched patterns
\citep{cornish:2004, aurich:2005, aurich:2006, then:2006, key:2007}.

In this work, we constrain the topology of the Universe using the method of 
matched circles proposed by \citet{cornish:1998} and apply it to
the 7-year \emph{WMAP} data \citep{jarosik:2010}. In contrast to
the majority of previous studies, we will pay special attention to the impact of
Galactic foreground residuals on the constraints. Some consideration
of this problem was made by \citet{then:2006} 
in their analysis of the first year \emph{WMAP} data release. We will
use also Monte Carlo (MC) simulations for the estimation of the false
detection level of the statistic. The 
method is applied to higher resolution maps than previously, which implies a
lower level of false detection and therefore tighter constraints on the size of the
Universe. As a result of computational limitations, we will restrict the 
analysis to a search for back-to-back circle pairs\footnote{pairs of
  circles centred around antipodal points}. 

In the following two sections we describe data used in analysis and
simulations of the CMB maps for a 
flat universe with the topology of a 3-torus which were used to test
the reliability of our implementation. The statistic adopted in our studies is
presented in \sect\ref{sec:statistic}. The results and conclusions are
described in the last two sections.

\section{Data} \label{sec:data}
The search for matched circles was performed on the 7-year \emph{WMAP}
data \citep{jarosik:2010}. The \emph{WMAP} satellite observes the sky
in five frequency bands denoted \emph{K}, \emph{Ka},  
\emph{Q}, \emph{V} and \emph{W}, centred on the frequencies of 22.8, 33.0, 40.7, 60.8, 93.5
GHz with angular resolutions of 52.8', 39.6', 30.6', 21' and 13.2', respectively. The
maps\footnote{available at http://lambda.gsfc.nasa.gov}  
are pixelized in the \textsc{healpix}\footnote{http://healpix.jpl.nasa.gov}
scheme \citep*{gorski:2005} with a resolution parameter
$N_{\rm side}=512$, corresponding to 3 145 728 pixels with a 
pixel size of $\sim 7$ arcmin.

In particular we used the Internal Linear Combination (ILC) map and the W-band map
corrected for Galactic emission outside of the applied mask using
a template fitting scheme \citep{gold:2010} and smoothed with a
gaussian beam profile of full width at half-maximum $({\rm FWHM}) = 20'$ to decrease noise
level. The former has been used for a number 
of full sky analyses, and versions available with earlier data releases have already been
exploited in the search for matched circles. The latter is the
highest angular resolution data measured by the 
\emph{WMAP} satellite. Since the false detection level for such a map is
lower, it provides the opportunity to estimate tighter constraints on topology.
We elected not to use data preprocessed as in \citet{cornish:2004}, i.e. 
a noise-weighted combination of the Q, V and W-band
maps smoothed to Q-band map resolution outside of the Kp2 Galactic
cut, and the ILC map within this region. Such a map has complex
properties since it consists of two regions with different angular
resolutions; moreover, as we will see the ILC map remains  
too contaminated by Galactic foreground residuals inside the Kp2 cut to
be used for cosmological analysis.

\section{Simulations of the CMB anisotropy maps in the multi-connected
  universe} \label{sec:simulations}
\indent To test the reliability of codes used to search for
the signature of a
multi-connected topology in the Universe, we performed simulations of
CMB skies for a flat universe with the topology of a 3-torus \citep{riazuelo:2004a}. The
simulations can be made in two ways: one can compute the covariance
matrix\footnote{assuming zero mean of the CMB maps} of the spherical
harmonic coefficients $a_{\ell m}$, $\mathbf{M}_{\ell m, (\ell m)'} \equiv \left<
a_{\ell m}^{} a_{\ell' m'}^\ast\right>$, \citep{riazuelo:2004b,phillips:2006} and then generate correlated $a_{\ell
m}$ coefficients using the Choleski decomposition of the covariance matrix
$\mathbf{M}=\mathbf{L}\, \mathbf{L}^\dag$ and uncorrelated unit
variance Gaussian variable $\boldsymbol{x}$, $\boldsymbol{a}=\mathbf{L}\,
\boldsymbol{x}$, or one can directly compute the $a_{\ell m}$ coefficients
using equation 
\begin{equation}
a_{\ell m}^T = - \frac{(2\pi)^3}{V} {\rm i}^\ell \sum_k \Delta_\ell^T(k,\Delta \tau) \frac{\sqrt{P(k)}}{k^2}
\sum_{|\boldsymbol{k}|=k} Y_{\ell m}^\ast(\Omega_{\boldsymbol{k}}) \hat{e}_{\boldsymbol{k}}
\end{equation}
where $\Delta \tau=\tau_0-\tau_r$ is difference between the present
conformal time $\tau_0$ and the time of 
recombination $\tau_r$, 
$P(k)$ is the density perturbation power spectrum, $\Delta_\ell^T
(k,\Delta \tau)$ is the response function for
temperature, $\hat{e}_{\boldsymbol{k}}$ is a Gaussian random variable which satisfies 
$\left<\hat{e}_{\boldsymbol{k}}
  \hat{e}_{\boldsymbol{k}'}^\ast\right>=\frac{V}{(2\pi)^3}
\delta_{\boldsymbol{k} \boldsymbol{k}'}$ and $V$ is the volume of the fundamental domain.
The sum over wavevectors $\boldsymbol{k}$ was split into two sums
$\sum_{\boldsymbol{k}}=\sum_k \sum_{|\boldsymbol{k}|=k}$ -- the sum over all possible
magnitudes of the $\boldsymbol{k}$ vector and over all $\boldsymbol{k}$ of magnitude $k$.

The advantage of the latter approach is a less aggressive scaling in
the computation time, $\ell_{\rm max}^5 V$ 
(number of $a_{\ell m}$ coefficients $\propto \ell_{\rm max}^2$ times
number of wave modes $\propto \ell_{\rm max}^3 V$) in comparison with the
former\footnote{the parity 
and symmetry relations can reduce by an order of magnitude the number of different elements in the matrix and 
computation time.} that goes as $\ell_{\rm max}^7 V$.
On the other hand, the disadvantage of the latter approach is that for each simulation we have to
repeat the computations\footnote{in the former case one need only
 generate a vector of Gaussian random numbers and multiply 
by the Choleski decomposed matrix. It is much faster than the full computation.}. Nevertheless, we 
decided to adopt this method since the difference in scaling is particularly
significant for the high resolution maps of interest in our studies. We also do not need more than one map for a
given topology to evaluate the possibility of detection.

The topology does not affect local physics, so the equations governing the evolution of cosmological 
perturbations are left unchanged. Thus, to compute the response functions, $\Delta_\ell^T (k,\Delta \tau)$,
one can use the publicly available \textsc{camb} code \citep{camb}. A change of
topology translates into a change of the modes that
can exist in the universe. Therefore, we need to compute the response function only for a set of allowed
wave numbers $k$. To speed up computations, we use in our code
precomputed response functions. Then, the most time consuming part is
the computation of the sum 
$\sum_{|\mathbf{k}|=k} Y_{\ell m}^\ast(\Omega_{\boldsymbol{k}}) \hat{e}_{\boldsymbol{k}}$.

The simulations do not include gravitational lensing of the CMB as the lensing deflection angle 
is small \citep{seljak:1996}, $\sim 3'$, compared to the minimal angular scale taken into account in 
the simulations, $\sim 20'$. The effect of the finite thickness of the last scattering surface is included in 
the simulations through the response function, $\Delta_\ell^T (k,\Delta \tau)$, computed by \textsc{camb}. However, 
the effect starts to be important at angular scales smaller than $\sim
4'$ and will therefore not be studied here.

\subsection{Requirements and numerical implementation}
To study the signatures of a given topology, a CMB map
is required with resolution comparable to the angular size of the beam profile
used to smooth the \emph{WMAP} W-band data (i.e.\ $\sim 20'$) to be analysed
here (see \sect\ref{sec:data}). We have adopted $\ell_{\rm max} = 500$
in our simulations. The dimension of the fundamental domain 
of the 3-torus was $L=2\, c/H_0$, which is about three times less than diameter of the
observable Universe i.e.\ $\sim 6.6\, c/H_0$. In such a
universe there are many pairs of matched circles of different
radii. The time needed for the generation of one such CMB map on a single
processor with clock speed 3 GHz is about 42 hours.

\section{Searching for the circles in the sky} \label{sec:statistic}
If light had sufficient time to cross the fundamental cell, an observer would see multiple copies of a single 
astronomical object. To have the best chance of seeing `around the universe' we should look for multiple 
images of the furthest reaches of the universe. Searching for multiple images of the last scattering surface -- 
the edge of the visible universe -- is then a powerful way to constrain
topology. Because the surface of last scattering is a sphere centred on the observer, each copy of the observer
will come with a copy of the last scattering surface, and if the copies are separated by a distance less than the 
diameter of the last scattering surface, then they will intersect along 
circles. These are visible by both copies of the observer, but from opposite sides. The two copies are really 
one observer so if space is sufficiently small, the CMB radiation from the last scattering surface will demonstrate
a pattern of hot and cold spots that match around the circles. 

The idea of using these circles to study topology
is due to \citet{cornish:1998}. Therein, a
statistical tool was developed to detect correlated circles in all sky maps of the CMB
anisotropy -- the circle comparison statistic
\begin{equation} \label{s_statistic}
S_{p,r}^{\pm} (\alpha, \phi_\ast)=\frac{\left<2\, T_p(\pm \phi) T_r(\phi+\phi_\ast)\right>}{\left<T_p(\phi)^2+T_r(\phi)^2\right>}\ ,
\end{equation}
where $\left< \right>=\int_0^{2\pi}d\phi$ and $T_p(\pm \phi)$, $T_r(\phi+\phi_\ast)$ are the temperature
fluctuations around two circles of angular radius $\alpha$  centered at
different points, $p$ and $r$, on the sky with relative phase $\phi_\ast$. The sign $\pm$ depends
on whether the points along both circles are ordered in a clockwise direction
(phased, sign $+$) or alternately whether along one of the circles the points are ordered in an
anti-clockwise direction (anti-phased, sign $-$). This allows the detection
of both orientable and non-orientable topologies.
For orientable topologies the matched circles have anti-phased
correlations while for non-orientable
topologies they have a mixture of anti-phased and phased correlations.
The statistic has a range over the interval $[-1,1]$. Circles that are 
perfectly matched have $S=1$, while uncorrelated circles will have a
mean value of $S=0$. To find matched circles for each radius $\alpha$, the
maximum value $S_{\rm max}^{\pm}(\alpha) = \underset{p,r,\phi_\ast}{\rm max}
\, S_{p,r}^{\pm}(\alpha,\phi_\ast)$ is determined.

In order to speed up the computations, one can use the fast Fourier transform (FFT) along the circles, 
$T_p(\phi) = \sum_m T_{p,m} \exp({\rm i}m\phi)$, and, by rewriting the statistic as 
$S_{p,r}^{+}(\alpha, \phi_\ast) = \sum_m s_m \exp(-{\rm i}m\phi_\ast)$, where 
$s_m = 2 \sum_m T_{p,m}^{} T_{r,m}^\ast / \sum_n \left( |T_{p,n}|^2+|T_{r,n}|^2\right)$, use 
inverse fast Fourier transform. It reduces the computational cost
$S_{p,r}^{\pm}(\alpha,\phi_\ast)$ from $\sim N$ to  $\sim N^{1/2} \log N$
operations, where $N$ is number of pixels in the map. 

In our studies we will use the statistic as modified in
\citet{cornish:2004} to include weighting of the $m$th harmonic of
the temperature around the $p$th circle $T_{p,m}$ by the factor $|m|$ that
takes into account the number of degrees of freedom per mode:
\begin{equation} \label{eqn:s_statistic_fft}
S_{p,r}^{+}(\alpha, \phi_\ast)=\frac{2 \sum_m |m| T_{p,m}^{} T_{r,m}^\ast
e^{-{\rm i} m \phi_\ast}}{\sum_n |n| \left( |T_{p,n}|^2+|T_{r,n}|^2\right)}\ ,
\end{equation}
where the Fourier coefficients $T_{p,m}$ are complex conjugated in the
case of the $S_{p,r}^{-}$ statistic. Such weighting enhances the
contribution of small scale structure relative to the large scales 
fluctuations. It is especially important because the large scale fluctuations 
which are dominated by the integrated Sachs-Wolfe (ISW) effect. This can obscure the 
image of the last scattering surface and reduce the ability to
recognise possible matched patterns on it.

The general search explores a six dimensional space: the location of the
first circle centre, $(\theta_p, \phi_p)$, the location of the second circle centre, $(\theta_r, \phi_r)$, the 
angular radius of the circle $\alpha$, and the relative phase of the two circles $\phi_\ast$. The number of 
operations needed to search for circle pairs scales as $\sim N^3 \log N$; thus 
implementation of a matched circles test is computationally very
intensive. Because of this we will restrict our analysis to a search for
pairs of circles centered around antipodal points, so called
back-to-back circles. Then, the search space can be reduced to four dimensions
and the number of operations to $\sim N^2 \log N$. We will not search for 
nearly back-to-back circles as in \citet{cornish:2004}, neither can we
apply their hierarchical approach to speed up computations.
In our case, we perform an analysis of high resolution maps ($\sim
20'$), so that the approach of initially degrading them to a lower
resolution and then refining the search in higher resolution maps only for
those circle pairs with the highest correlations risks missing some pairs of
circles. As a consequence, our analysis is more computationally demanding
than the analysis done by \citet{cornish:2004} preventing the study of
a more general class of topologies. Although the back-to-back search can 
detect a large class of the topologies that we might hope to find, there remain
those topologies that predict matched circles without antipodal configurations such as the 
Hantzsche-Wendt space for flat universes or the Picard space for hyperbolic universes 
\citep{aurich:2004}.

We used the \textsc{healpix} scheme with resolution
parameter $N_{\rm side} = 512$ to define our search grid on the
sky. To accommodate the use of the FFT approach,
$n=2^{r+1}$ points are used for each circle, where $r$ corresponds to resolution
parameter of the map given by $N_{\rm side} = 2^r$. By definition this is
also the angular resolution used to step through $\phi_{\ast}$. For $\alpha$, we used steps
a little bit smaller than characteristic scale $\theta_c$ of
coherent structures in the map, i.e.\ $2\, \theta_c / 3$. The scale of coherence
was approximated by the Full Width at Half Maximum (FWHM) of 
a gaussian beam profile used to smooth maps. The values of the temperature anisotropy at each
point along the circle were interpolated based on values for the four
nearest pixels with weights inversely proportional to the distances between
a given point and the pixel centers. Other methods of interpolation
can be used but we have verified that even using the temperature value
of the nearest pixel does not change the value of the statistic significantly.

To draw any conclusions from an analysis based on the statistic $S_{\rm
 max}^{\pm}(\alpha)$ it is very important to correctly estimate the threshold
for a statistically significant match of the circle pairs. The chance of random
matches is inversely proportional to the number of coherent structures
along the circles, therefore a false positive signal level of the statistic is
especially large for circles with smaller radii. We
used simulations of the maps with the same noise properties and
smoothing scales as the \emph{WMAP} data to establish the threshold
such that fewer than 1 in
100 simulations would yield a false event. 

It is important to note that the false detection level is smaller for
higher resolution maps. Conversely, as shown by \citet{cornish:1998},
the value of the
statistic for matched circles 
\begin{equation} \label{}
S_{\rm max} \approx {\xi^2 \over 1+\xi^2} \ ,
\end{equation}
depends on the ratio of the signal rms $\sigma_s$ to the noise rms
$\sigma_n$ ratio, denoted $\xi$ for the map, 
thus the efficiency of the statistic to detect matched circles is increased by smoothing the data. 
A smoothing scale should therefore be adopted that is a reasonable
trade-off between these two requirements. 

To eliminate the regions most contaminated by Galactic foreground
residuals, we utilise the masks defined by the \emph{WMAP} team. However,
the statistic is very sensitive to the masking, particularly if a
significant fraction of one or both circle pairs 
lies in the masked region. In this case, there is a significant probability
to find a chance correlation of the temperature pattern between unmasked parts of the circles,
thus increasing the false detection level. The effect is the most
pronounced for circles with the largest radii, close to 90 degrees, as well
with very small radii. To avoid this, we restrict our analysis
to those pairs for which less than half the length of each circle
is masked. Though, the statistic computed in this way is not optimal,
it is much more robust with respect to masking.
It should be noted that the dependence of the statistic on the cut could be avoided
if the statistic were expressed as a function of the number of coherent structures
along the circles. However, because our principal goal is to constrain the size of the 
Universe it is better to use the statistic determined as a function of
circle radius.

Finally, \cite{key:2007}  have suggested that the performance of the statistic
(\ref{eqn:s_statistic_fft}) may be further improved by bandpass
filtering the input map to minimise the anisotropy contribution from the ISW and Doppler
terms which blur any signatures of topology present on the last
scattering surface. We chose not to apply this kind of filtering because of
possible complications arising from the interaction of the mask and the
filter. Our analysis is enhanced by the use of high resolution maps
rather than filtering.

\section{Results} \label{sec:results}

\begin{figure*}

\centerline{
\epsfig{file=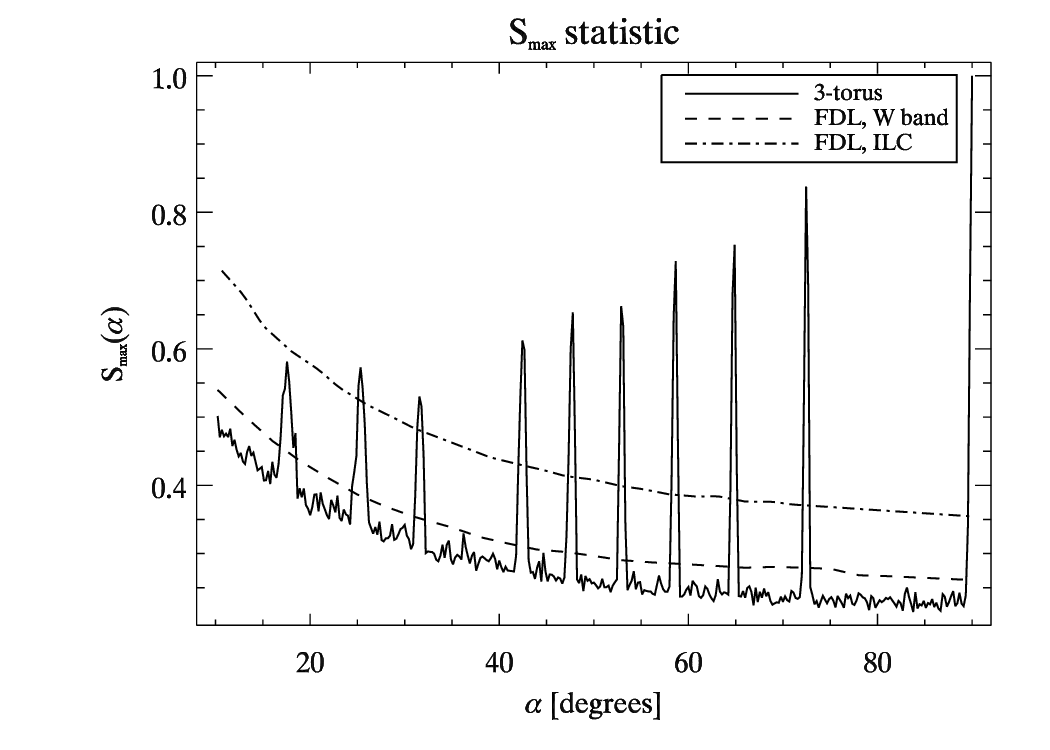,scale=0.75}
}

\caption{An example of the $S_{\rm max}^{-}$ statistic for a simulated CMB map of
  universe with the topology of a cubic 3-torus with dimensions $L = 2\ c/
  H_0$. The dashed and dot-dashed lines show the false detection level for the statistic
  estimated from 100 MC simulations of the \emph{WMAP} 7-year W-band
  map (the same as in Fig.\ 3) and ILC map (the same as in Fig.\ 2), respectively.}
\label{fig:smax_t222}
\end{figure*}

Before beginning the search for pairs of matched circles in the
\emph{WMAP} data, we
validated our algorithm using simulations of the CMB sky
for a universe with 3-torus topology for which the dimension of the cubic fundamental domain
$L=2\, c/H_0$, and with cosmological parameters corresponding
to the $\Lambda CDM$ model \citep[see][Table 3]{larson:2010} determined from the
7-year \emph{WMAP} results. In particular, we verified that our code is able to find all pairs of
matched circles in such map. The statistic $S_{\rm max}^{-}(\alpha)$
for the map is shown in \fig\ref{fig:smax_t222}. Note that the
peak amplitudes in the statistic, corresponding to the temperature correlation for
matched circles, decrease with radius of the circles. \citet{cornish:2004},
noted that this is primarily caused by the Doppler term
which becomes increasingly anticorrelated for circles with radius
smaller than $45^\circ$. The temperature match on the last
scattering surface will be also diluted by the ISW effect which comes
from the evolution of the structures close to the observer. 
However, to a large extent the $m$ weighting used in the
statistic (\ref{eqn:s_statistic_fft}) mitigates against this and
prevents the statistic from being dominated by the large scales.

\begin{figure*}

\centerline{
\epsfig{file=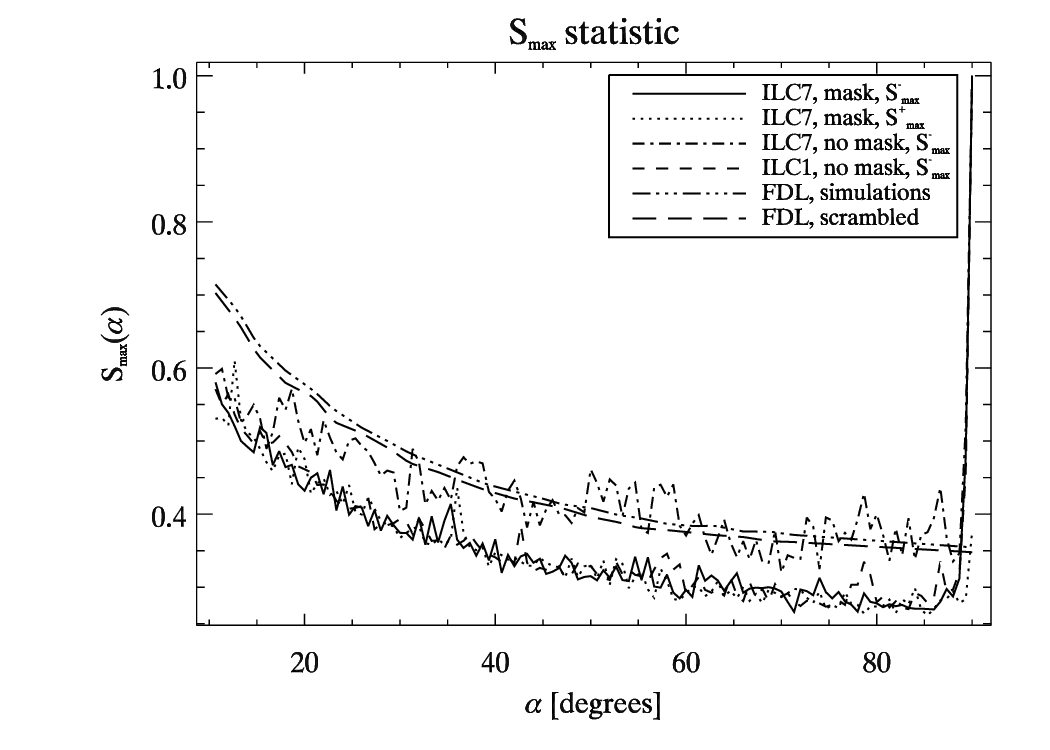,scale=0.75}
}

\caption{$S_{\rm max}^{\pm}$ statistic for the \emph{WMAP} 7-year ILC
  map. The solid and dotted lines show the statistics $S_{\rm max}^-$ 
  and $S_{\rm max}^+$, respectively, for the ILC map masked with the
  KQ85y7 mask. The dash-dotted and dashed 
  lines show the statistic $S_{\rm max}^-$ for the 7-year and the
  first year ILC unmasked maps, respectively. The false detection
  level  estimated from 100 MC simulations of the ILC map and by scrambling the $a_{\ell m}$
  coefficients of the map are marked by dash-three dots and long dashed
  lines, respectively. The peak at $90^\circ$ corresponds to a match
  between two copies of the same circle of radius $90^\circ$ centered
  around two antipodal points.}
\label{fig:smax_wmap_ilc}
\end{figure*}

\begin{figure*}

\centerline{
\epsfig{file=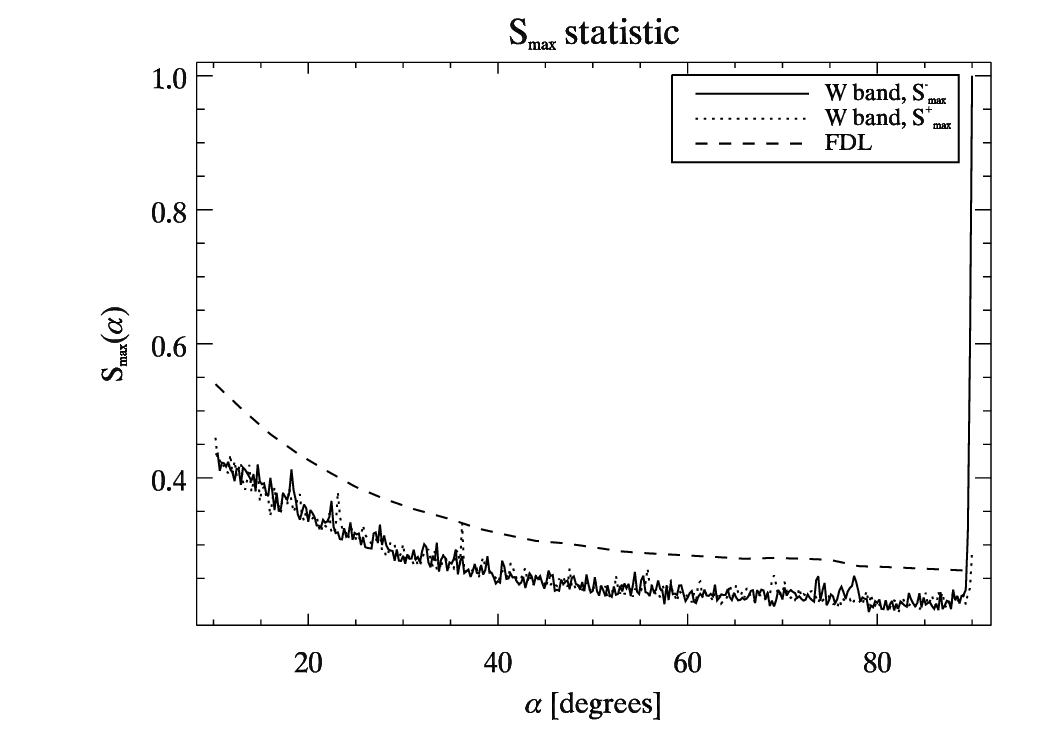,scale=0.75}
}

\caption{$S_{\rm max}^{\pm}$ statistic for the \emph{WMAP} 7-year W-band
  map. Solid and dotted lines show the statistics $S_{\rm max}^-$ and
  $S_{\rm max}^+$, respectively, for the W-band map masked with the
  KQ85y7 mask. The dashed line  
  is the false detection level estimated from 100 MC simulations.
  The peak at $90^\circ$ corresponds to a match
  between two copies of the same circle of radius $90^\circ$ centered
  around two antipodal points.}
\label{fig:smax_wmap_w}
\end{figure*}

The intersection of the peaks in the matching statistic with the false 
detection level estimated for the W-band data 
(the same as in \fig\ref{fig:smax_wmap_w}) defines the minimum radius of the correlated circles which
can be detected for this map. The height of the peak with the smallest
radius seen in \fig\ref{fig:smax_t222} indicates that the minimum
radius is about $\alpha_{\rm min} \approx 10^\circ$. To emphasise the advantage
of using higher resolution data in the analysis, we also show the false
detection level estimated for the ILC map (the same as in
\fig\ref{fig:smax_wmap_ilc}). In this case,
the minimum radius for detectable matched circles
is about $\alpha_{\rm min} \approx 25^\circ$. 

The statistics for the \emph{WMAP} ILC map are shown in
\fig\ref{fig:smax_wmap_ilc}. The map was analyzed on both the full sky
and after applying the KQ85y7 mask \citep{gold:2010}.
In the case of the full sky analysis, the $S_{\rm max}^-$ statistic shows some excess
for pairs of circle with large radii. Nevertheless, after removing the most
foreground contaminated regions the statistic does not reveal any
unusually large values. This indicates that residuals of the Galactic
emission, especially in the Galactic plane, cannot be neglected in the
analysis of matched circles.

It is interesting that similar behaviour is not seen for a full sky analysis of the first
year \emph{WMAP} ILC map. This probably explains why the importance of masking
the most contaminated regions was not recognised in previous papers
using the matched circles statistic, 
the only exception being the paper by \citet{then:2006}. However, because
they studied only the first year \emph{WMAP} data, they also came to the
conclusion that the ILC map is good enough for the full sky analysis. 
We checked that the excess is not related to
the bias correction (for residual foreground emission) applied to the
7-year ILC map by the \emph{WMAP} team
since this was not implemented
for the first year data. Indeed, the excess remains present for the 7-year ILC
constructed by simply coadding the individual
frequency channels with the weights provided in \citet{gold:2010}. 
Deeper studies of this problem are beyond the scope
of this paper. Nevertheless, it provides a further warning against the naive use of a
full sky ILC map for cosmological analysis, and particularly with
respect to the two-point correlation function which is closely related to the statistic used
here\footnote{The remarks of \cite{efstathiou:2010} regarding the lack
  of evidence for Galactic foreground residuals at low Galactic
  latitudes are only relevant for the ILC map when smoothed to
  $10^\circ$ resolution.}.

The false detection level shown in \fig\ref{fig:smax_wmap_ilc}
was estimated on the basis of 100 simulations of the ILC map with the KQ85y7 mask
applied, assuming a simple-connected universe and established from the
requirement that fewer 
than 1 in 100 simulations should yield a false match. We did not correct the
simulated maps for the bias coming from the random correlation between the CMB and
the Galactic foreground\footnote{so called `Cosmic Covariance'},
since the effect is small outside of the Galactic plane region.
Moreover, such a correction does not appear to be 
reliably evaluated as demonstrated with simulations that have been generated
using the same foreground templates as for the simulated ILC maps. 
Maps corrected in this way do not
properly reflect uncertainties concerning details of the Galactic emission,
especially in the Galactic plane. 

For comparison we show also the false detection level estimated on the
basis of a `scrambled' version of the ILC map (i.e. by randomly exchanging the
$m$ spherical harmonic coefficients, $a_{\ell m}$, of the map at every $\ell$) as in
\citet{cornish:2004}. However, we still applied the KQ85y7 mask 
for each `scrambled' map to be consistent with simulations of
the ILC data. As can be seen, the false detection level is slightly lower in this case. It is not surprising
because the scrambling generates maps with the same two-point function
and only different phase correlations that results in a smaller variance of the
$S$-statistic than for an ensemble of simulated ILC maps.

In order to decrease the false detection level and be able to detect
matched circles with smaller radius, we analyzed 
also the \emph{WMAP} data with the highest angular resolution i.e.\ the W-band
map, corrected for Galactic foregrounds and smoothed with a gaussian
beam profile of ${\rm FWHM} = 20'$ to decrease the noise level. 
The statistic for this map analysed with the KQ85y7 mask is shown in
\fig\ref{fig:smax_wmap_w}. As for the ILC map, the false detection
level was estimated from 100 simulations of the W-band data.

We did not find any statistically significant correlation of circle
pairs in either map. As seen in \fig\ref{fig:smax_t222}, the
minimum radius at which the peaks expected for the matching statistic are larger than
the false detection level is about $\alpha_{\rm min} \approx
10^\circ$ for the W-band map. Thus, we can exclude any topology that predicts matching 
pairs of back-to-back circles larger than this radius. This implies
that in a flat universe described otherwise by the best-fit 7-year \emph{WMAP}
cosmological parameters, a lower bound on
the size of the fundamental domain is $d = 2\, R_{\rm LSS} \cos(\alpha_{\rm min})
\simeq 27.9\ \rm{Gpc}$, where $R_{\rm LSS}$ is the distance to the last
scattering surface. 

Of course, the above constraints do not apply to those universes for
which the orientation of the matched circles is impossible to detect
due to partial masking on the sky. Examples of topologies which
can be overlooked are so called slab and chimney spaces
\citep{riazuelo:2004a}. For such topologies and appropriate configuration, all pairs of
matched circles could lie in the Galactic plane that is removed by the
mask. The probability of overlooking the circles depends on the specific
topology and radii of the circles, so it is difficult to give a
general expression. Nevertheless, one can suppose that it 
decreases as the fraction of the masked sky decreases.

\section{Conclusions} \label{sec:conclusions}
We have studied constraints on the topology of the Universe using the
method of matched circles as applied to the 7-year \emph{WMAP} ILC map 
and the foreground-reduced W-band map. We paid special attention to
three aspects of the analysis that have been neglected in previous
studies --
the application of a mask, the use of high resolution data and the estimation
of the false detection level on the basis of detailed MC simulations of the sky
maps.

The necessity for the application of a mask is due to the presence of
residual Galactic foreground emission present even in the ILC map. 
These introduce a non-negligible effect 
on the matched circles statistic that is used for constraining
topology. However, the possibility to apply the analysis to masked
maps yields the opportunity
to more tightly constrain topology by using higher resolution,
foreground corrected \emph{WMAP} W-band data. 
Constraints on the topology depend significantly on the threshold
for a significant match of the 
circle pairs. In order to estimate this correctly, we used 100 MC simulations of 
the ILC and W-band maps assuming a simply-connected universe.
The level of false detection calibrated in this 
manner is slightly higher than that derived in \citet{cornish:2004}
using a method in which the analyzed maps are resampled 
by shuffling their spherical harmonic coefficients.
Although the difference is not very big, it is
worth noting that the constraints on the size of Universe are
overestimated if one uses a lower level of false detection.

The analysis of the \emph{WMAP} W-band map, after correction using
templates of Galactic foreground emission, did not reveal any significant
correlations for pairs of back-to-back circles with a radius greater
than $\sim 10^\circ$. This substantially extends the previous
constraint on the 
minimum radius of detectable matched circles given by
\citet{key:2007} of $20^\circ$. It also places a lower  
bound on the size of fundamental domain of about 27.9 Gpc for a flat
universe described by the best-fit 7-year \emph{WMAP} cosmological parameters.
Although this constraint concerns only those
universes with such dimensions and orientation of the fundamental domain
with respect to the mask that allow the detection of pairs of matched circles,
the probability of overlooking circle pairs is rather low
for the KQ85y7 mask that removes only a relatively small fraction of the
sky. 

Of course, observations of the CMB with higher angular resolution and 
significantly lower noise level by the \emph{Planck}
satellite may yield even tighter constraints on the topology of the
Universe. However,
one should bear in mind that the possible improvement in the lower
bound on the size of the Universe will not be substantial. The
current constraint is not much smaller than the
diameter of the observable Universe $2\, R_{\rm LSS} = 28.3\ {\rm Gpc}$,
which imposes an upper limit on the size of the fundamental domain 
that it is possible to detect using the method of matched
circles. The only significant improvement might be related to improved
modeling of the Galactic
emission allowing the application of a smaller mask approaching 
closer to the Galactic
plane. This would minimize the probability of overlooking topologies with
matched circles that are hidden within the masked region of the sky.

Finally, as in \citet{cornish:2004}, the studies could also be
extended to search for nearly back-to-back 
circle pairs. However, the much higher computational requirements for
the analysis of high resolution maps make such studies difficult and extremely
time-consuming. Nevertheless, we can hope that the steadily increasing 
speed of processors and availability of larger computational resources
will make such computations feasible in the coming years, thus
allowing a final resolution of the problem of the topology of our Universe.

\section*{Acknowledgments}
We acknowledge use of \textsc{camb}\footnote{http://camb.info/} \citep{camb}
and the \textsc{healpix} software
\citep{gorski:2005} and analysis package for deriving the results in
this paper. We acknowledge the use of the Legacy Archive for
Microwave Background Data Analysis
(LAMBDA). Support for LAMBDA is provided by the NASA Office of Space
Science. The authors acknowledge the use of version 1.6.6 of the Planck
Sky Model, developed by the Component Separation Working Group (WG2)
of the \emph{Planck} Collaboration \citep{leach:2008,betoule:2009}. Part of the computations were
performed at Interdisciplinary Center for Mathematical and
Computational Modeling at Warsaw University within a grant of
computing time no.\ G27-13. This research was
supported by the Agence Nationale de Recherche (ANR-08-CEXC-0002-01).

%------------------------------------


\begin{thebibliography}{}

\bibitem[\protect\citeauthoryear{Aurich et al.}{2004}]{aurich:2004} 
Aurich R., Lustig S., Steiner F., Then H., 2004, CQGra, 21, 4901 

\bibitem[\protect\citeauthoryear{Aurich, Lustig, \& Steiner}{2005}]{aurich:2005} 
Aurich R., Lustig S., Steiner F., 2005, CQGra, 22, 2061 

\bibitem[\protect\citeauthoryear{Aurich, Lustig, \& Steiner}{2006}]{aurich:2006} 
Aurich R., Lustig S., Steiner F., 2006, MNRAS, 369, 240 

\bibitem[\protect\citeauthoryear{Betoule et al.}{2009}]{betoule:2009} 
Betoule M., Pierpaoli E., Delabrouille J., Le Jeune M., Cardoso J.-F., 2009, A\&A, 503, 691 

\bibitem[Copi et al.(2004)]{copi:2004} 
Copi C. J., Huterer D., \& Starkman G. D., 2004, Phys. Rev. D., 70, 043515
%%CITATION = ASTRO-PH 0310511;%%

\bibitem[\protect\citeauthoryear{Cornish, Spergel, \& Starkman}{1998}]{cornish:1998} 
Cornish N.~J., Spergel D.~N., Starkman G.~D., 1998, CQGra, 15, 2657 

\bibitem[\protect\citeauthoryear{Cornish et al.}{2004}]{cornish:2004} 
Cornish N.~J., Spergel D.~N., Starkman G.~D., Komatsu E., 2004, PhRvL, 92, 201302 


\bibitem[de Oliveira-Costa et al.(2004)]{de Oliveira-Costa:2004}
de Oliveira-Costa A., Tegmark M., Zaldarriaga M., Hamilton
A. 2004, Phys. Rev. D., 69, 063516
%%CITATION = ASTRO-PH 0307282;%%

\bibitem[\protect\citeauthoryear{Efstathiou, Ma, \& Hanson}{2010}]{efstathiou:2010} 
Efstathiou G., Ma Y.-Z., Hanson D., 2010, MNRAS, 407, 2530 


\bibitem[Eriksen et al.(2004)]{eriksen:2004} 
Eriksen H. K., Hansen F. K., Banday A. J., G{\'o}rski K. M., \&
Lilje P. B., 2004, ApJ, 605, 14 
%%CITATION = ASTRO-PH 0307507;%%

\bibitem[\protect\citeauthoryear{Gold et al.}{2010}]{gold:2010} 
Gold B., et al., 2010, arXiv, arXiv:1001.4555 

\bibitem[G{\'o}rski et al.(2005)]{gorski:2005} 
G{\' o}rski K.~M., Hivon E., Banday A.~J., Wandelt B.~D., Hansen F.~K., Reinecke M., \&
Bartelmann M., 2005, ApJ, 622, 759 
%%CITATION = ASTRO-PH 0409513;%%

\bibitem[Hansen et al.(2004)]{hansen:2004} 
Hansen F. K., Banday A.~J., \& G{\' o}rski K.~M. 2004, 
MNRAS, 354, 641
%%CITATION = ASTRO-PH 0404206;%%

\bibitem[\protect\citeauthoryear{Jarosik et al.}{2010}]{jarosik:2010} 
Jarosik N., et al., 2010, arXiv, arXiv:1001.4744 

\bibitem[\protect\citeauthoryear{Key et al.}{2007}]{key:2007} 
Key J.~S., Cornish N.~J., Spergel D.~N., Starkman G.~D., 2007, PhRvD,
75, 084034

\bibitem[\protect\citeauthoryear{Kunz et al.}{2006}]{kunz:2006} 
Kunz M., Aghanim N., Cayon L., Forni O., Riazuelo A., Uzan J.~P., 2006, PhRvD, 73, 023511 

\bibitem[\protect\citeauthoryear{Kunz et al.}{2008}]{kunz:2008} 
Kunz M., Aghanim N., Riazuelo A., Forni O., 2008, PhRvD, 77, 023525 

\bibitem[\protect\citeauthoryear{Larson et al.}{2010}]{larson:2010} 
Larson D., et al., 2010, arXiv, arXiv:1001.4635 

\bibitem[\protect\citeauthoryear{Leach et al.}{2008}]{leach:2008} 
Leach S.~M., et al., 2008, A\&A, 491, 597 

\bibitem[\protect\citeauthoryear{Lewis, Challinor, \& Lasenby}{2000}]{camb} 
Lewis A., Challinor A., Lasenby A., 2000, ApJ, 538, 473 

\bibitem[\protect\citeauthoryear{Phillips \& Kogut}{2006}]{phillips:2006} 
Phillips N.~G., Kogut A., 2006, ApJ, 645, 820 


\bibitem[\protect\citeauthoryear{Riazuelo et al.}{2004a}]{riazuelo:2004a} 
Riazuelo A., Uzan J.-P., Lehoucq R., Weeks J., 2004a, PhRvD, 69, 103514 

\bibitem[\protect\citeauthoryear{Riazuelo et al.}{2004b}]{riazuelo:2004b} 
Riazuelo A., Weeks J., Uzan J.-P., Lehoucq R., Luminet J.-P., 2004b, PhRvD, 69, 103518 

\bibitem[Schwarz et al.(2004)]{schwarz:2004} 
Schwarz D.~J., Starkman G.~D., Huterer D., Copi, C.~J., 2004, Phys.~Rev.~Lett., 93, 221301
%%CITATION = ASTRO-PH 0403353;%%

\bibitem[Seljak (1996)]{seljak:1996} Seljak U., 1996, ApJ, 463, 1

\bibitem[\protect\citeauthoryear{Then}{2006}]{then:2006} 
Then H., 2006, MNRAS, 373, 139 

\end{thebibliography}
\end{document}